\documentclass[prl,superscriptaddress,twocolumn,amsmath,floatfix]{revtex4}
\usepackage{graphicx}

\newcommand{\NII}{{
    National Institute of Informatics,
    2-1-2 Hitotsubashi, Chiyoda-ku,
    Tokyo 101-8430, Japan
}}

\newcommand{\Ginzton}{{
    Edward L. Ginzton Laboratory,
    Stanford University,
    Stanford, California 94305-4088, USA
}}

\newcommand{\Paderborn}{{
    Dept. of Physics,
    University of Paderborn,
    Warburger Str. 100, 33098
    Paderborn, Germany
}}

\newcommand{\ts }[2]{#1_{\text{#2}}}          
\newcommand{\tsc}[2]{#1_{\text{\textsc{#2}}}} 
\newcommand{\ket}[1]{\ensuremath{|#1\rangle}} 

\begin{document}

\title{Indistinguishable photons from independent semiconductor single-photon devices}
    \author{Kaoru Sanaka}
    \email[email:]{sanaka@stanford.edu}
    \affiliation{\Ginzton} \affiliation{\NII}
    \author{Alexander Pawlis}
    \affiliation{\Ginzton} \affiliation\Paderborn
    \author{Thaddeus D. Ladd}
    \affiliation{\Ginzton} \affiliation{\NII}
    \author{Klaus Lischka}
    \affiliation\Paderborn
    \author{Yoshihisa Yamamoto}
    \affiliation\Ginzton\affiliation\NII
\date\today

\begin{abstract}

We demonstrate quantum interference between photons generated
by the radiative decay processes of excitons that are bound to
isolated fluorine donor impurities in ZnSe/ZnMgSe quantum-well
nanostructures.  The ability to generate single photons from these devices is confirmed by auto-correlation experiments, and indistinguishability of single photons from two independent devices is confirmed via a Hong-Ou-Mandel dip. These results indicate that donor impurities in appropriately engineered semiconductor structures can portray atom-like homogeneity and coherence properties, potentially enabling scalable technologies for future large-scale optical quantum
computers and quantum communication networks.
\end{abstract}

\maketitle

Many schemes for scalable optical information processing rely
critically on the quantum interference of single photons
generated by a large number of independent triggered single photon
sources, or on highly homogeneous interactions between many photons
and many matter qubits~\cite{cy95,LOQC_papers,HOM_entanglement_papers,communication_papers}.  Sufficient
homogeneity is certainly available in trapped atoms~\cite{indistinguishable_atoms} and
ions~\cite{indistinguishable_ions,Monroe_teleportation},  but the
scalability of such systems is challenging, in large part due to the
requirements of laser cooling and trapping.   On the other hand,
solid-state quantum computers and quantum communication networks
pose a different challenge: finding materials in which many
independent atom-like levels efficiently generate identical single
photons or scatter single photons identically.  For example, quantum
computation using only linear optics is possible~\cite{LOQC_papers},
and all of the optical components may be implemented on a
chip~\cite{LOQC_chip}, but a remaining required resource so far
lacking is a large array of triggered, independent sources of
indistinguishable single photons.

\begin{figure}
\includegraphics[width=0.9\columnwidth]{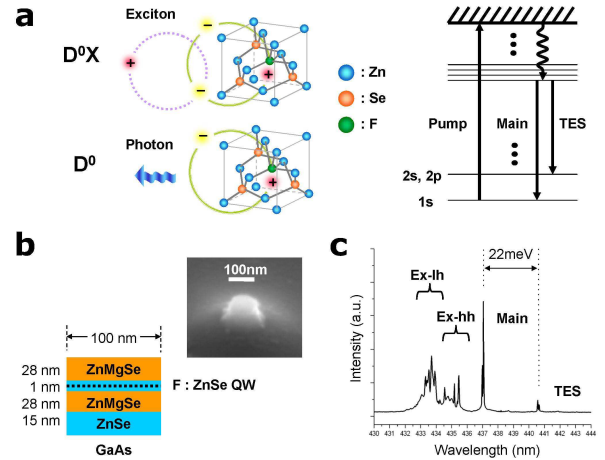}
\caption{\label{devicespectrum} (Color online) (a)  Schematic of a
neutral fluorine donor in ZnSe (D$^0$) and the associated
donor-bound exciton (D$^0$X) state, and corresponding energy diagram
showing the above-band pump, the D$^0$X-to-D$^0$ transition (Main),
and the two-electron-satellite (TES).  (b) The material composition
of the single-photon device grown by molecular beam epitaxy, and a
scanning electron micrograph of one device structure.
(c) Emission spectra of a typical device, showing
photoluminescence from the Main and TES transitions indicated by the
energy diagram, as well as light-hole free excitons (FX-lh) and
heavy-hole free excitons (FX-hh). }
\end{figure}

Single photons sequentially generated from one semiconductor source,
such as a self-assembled InAs quantum dot (QD) in a
distributed-Bragg-reflector (DBR)~\cite{sf02} or photonic crystal
(PC)~\cite{efzyv07} microcavity, have been used to demonstrate
linear optics quantum computing concepts, for example a quantum
teleportation gate~\cite{sf04b}.  Such sources also show another
important advantage of a semiconductor system: the possibility of
electrical pumping~\cite{electrical_pump_qdot}. However, quantum
interference between photons emitted by independent QDs has not yet
been realized, principally because self-assembled QDs are not
identical and consequently show a broad distribution of emission
wavelengths.

Here, we experimentally demonstrate indistinguishability of independent semiconductor sources based on the radiative
recombination of excitons bound to neutral donors isolated by
semiconductor nanostructures.    In particular, we use the fluorine
donor in ZnSe/ZnMgSe quantum wells (QWs)~\cite{Pawlis2006,ZnSelasing}.


Isolated impurity-bound excitons are attractive as single-photon
emitters since the impurity-bound-exciton-related emission has a
well defined emission wavelength with a small inhomogeneous
linewidth.  Further, the single electron spin of the neutral
donor in its orbital ground state can be used as a long-lived matter
qubit.  Donors are better suited to such applications than acceptors~\cite{Strauf2002} or isoelectronic impurities~\cite{isoelectronic_ZnSe}, since acceptor-bound-holes suffer from the rapid, spin-orbit-related relaxation times and isoelectronic impurities do not have the needed metastable ground states.  Several proposals~\cite{HOM_entanglement_papers} and a recent demonstration with trapped ions~\cite{Monroe_teleportation} demonstrate that indistinguishable photon emission into two orthogonal modes, in this case corresponding to emission from the neutral donor-bound exciton state (D$^0$X) to the two electron spin states of a neutral donor (D$^0$), may entangle distant spins heralded by an unbunched, two-photon coincidence counts; the degree to which these schemes are robust to interferometric stability and timing jitter is indicated by the width of the Hong-Ou-Mandel (HOM) dip, which we measure here.  The electron spin states might also be interfaced with a photonic quantum
communication bus through cavity-enhancement of the transitions following a variety of proposals~\cite{communication_papers}. These proposals require high homogeneity in the optical transitions, which is also prerequisite for the HOM dip.  Our demonstration therefore suggests that this system may be useful as an optically addressable quantum memory in future quantum repeaters for long-distance quantum communication or in quantum computers with chip-based photonic ``wiring."

\begin{figure}
\includegraphics[width=0.9\columnwidth]{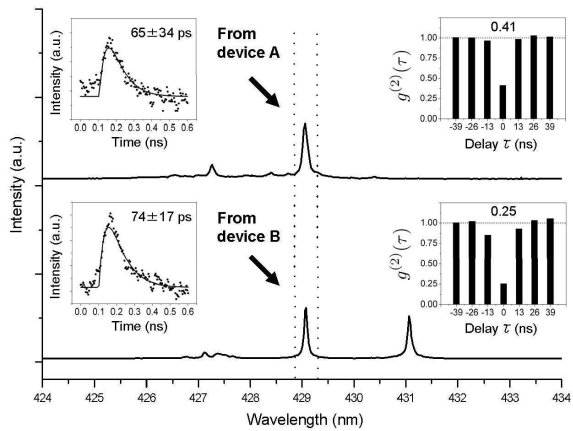}
\caption{\label{spectra} Emission spectra of device A and device B
under pulsed above-bandgap (410 nm) excitation.
The arrows indicate emission lines from two different devices; the linewidth seen here is the 0.01~nm resolution limit of the spectrometer.  The dotted lines indicate the spectral window used when collecting photon statistics.
The right insets show the normalized photon correlation histogram of the emission lines noted by arrows. The left insets are the
time-dependent emission decay functions as measured by a streak
camera, with 5-point running averaging and an exponential fit (solid
line). }
\end{figure}

The fluorine donor in ZnSe, depicted in Fig.~\ref{devicespectrum}a,
is particularly well-suited for such applications, since nuclear
decoherence of the electron spin may be suppressed by isotopic
purification~\cite{ZnSe_isotopicgrowth,thewalt}, unlike in
\textsc{iii-v} systems.  Moreover, the spin-1/2 $^{19}$F nucleus with
100$\%$ abundance provides even more potential for long-lived
quantum memory, since electron entanglement could be transferred to this longer-lived spin via double resonance techniques while building larger entangled states~\cite{benjamin_brokered}.  At the same time, this system can be engineered using many of the sophisticated techniques of modern nanotechnology employed with \textsc{iii-v} semiconductor alloy systems.  Heterostructures and quantum wells (QWs) based on wide bandgap  \textsc{ii-vi} semiconductor alloys (ZnSe, MgSe, ZnS, MgSe) may be  grown nearly defect-free on GaAs substrates~\cite{ZnSe1998}.
Microdisk cavities and waveguide-structures, for example, can be
realized by combinations of dry-etching processes of ZnSe with
selective wet-chemical etching of the GaAs
substrate~\cite{ZnSe_wetetch}.

A detailed study of the properties of the fluorine donor in
ZnSe/ZnMgSe QWs was performed in Ref.~\onlinecite{Pawlis2006} and
recently, low-threshold microcavity lasers have been realized with
similar $\delta$-doped QW structures in Ref.~\onlinecite{ZnSelasing}. For the present study, the total multilayer structure shown in Fig.~\ref{devicespectrum}b was grown on a (100)-GaAs substrate and a 15 nm buffer layer of ZnSe, to guarantee optimal interface properties. The ZnSe QW has a thickness of 1~nm and is sandwiched between two 28-nm-thick ZnMgSe barrier layers with a magnesium concentration of about 13$\%$. The fluorine $\delta$-doping in the central region of the ZnSe QW was performed with a sheet donor concentration of approximately $3\times 10^{10}$~cm$^{-2}$. The nanofabrication of posts with 100~nm
diameter was done with electron-beam writing and wet-chemical
etching~\cite{ZnSe_wetetch}; a scanning-electron-micrograph of one
such post is also shown in Fig.~\ref{devicespectrum}b. In the
100-nm-diameter post-nanostructures, there are only 2.4 fluorine
donor atoms on average, which may be separately observed due to a
small strain-induced inhomogeneous broadening.

Figure~\ref{devicespectrum}c shows the spectrum of a single
donor-bound exciton emitted by a typical device. The excited levels
of the D$^0$X complex correspond to different rotational states of
the bound hole, but these states are nearly degenerate in zero
magnetic field. Although the D$^0$X state predominantly relaxes to
the 1s D$^0$ state, there is a finite probability that the state
relaxes to any of the excited (2s, 2p, etc.) D$^0$ states, predicted
by a simple hydrogen-like model. These two-electron satellite (TES)
transitions can be observed about 22 meV below the main line, and
serve to verify that these transitions are due to
neutral-donor-bound excitons.

From many devices of varying properties, we have chosen two for this
study, denoted A and B, with spectra shown in Fig.~\ref{spectra}.
The devices are excited above the bandgap using a 3~ps pulse from a
frequency-doubled Ti:sapphire mode-locked laser.  The spectra from these particular devices are shifted to shorter wavelengths than those in Fig.~\ref{devicespectrum}c due to slightly different magnesium content and strain status in the chip. Both devices emit at the same wavelength of 429.06~nm  within the 0.01~nm resolution limit of the spectrometer.  By fitting the spectra to Lorentzian lineshapes, the two center wavelengths are found to be identical within an accuracy of 1.5~GHz.  Although not all devices we observe are as homogeneous as this pair, the probability of finding such a pair vastly exceeds that for QDs. This probability could be further enhanced with small amounts of local tuning, which may be possible via local heating or Stark shifting~\cite{lettow07}. The right insets of Fig. 2 show the normalized number of coincident photon counts versus delay $\tau$ between two photodetectors following a 50/50 beam-splitter [$g^{(2)}(\tau)$]. This photon correlation histogram features a series of peaks separated by 13 ns, the repetition period of the pulsed laser.  (The shape of each peak would be determined by the detector timing resolution, but we have integrated over this shape).  Due mostly to photon collection inefficiency, one photon is counted from each device for every 10$^{4}$ pump pulses.  The residual two-photon probability at zero delay is measured to be $g^{(2)}(0)$ = 0.41 and 0.25 for devices A and B.  This certainly indicates the emission of sub-Poissonian light from these
devices (unlike laser or thermal sources), but there remain residual
two-photon counts due to emission from the tail of the free-exciton
spectrum as well as residual light from nearby devices.  There is also slight antibunching evident at $\tau=\pm 13$~ns, likely due to a longer-lived nearby charge fluctuator~\cite{charlieprb}.

Also in the left insets of Fig.~\ref{spectra} are the time-dependent
decay functions for each device as measured by a streak camera,
showing that each emitter has a lifetime of less than 100~ps. This
is comparable to the lifetime of InAs QD excitons shortened by a DBR
microcavity structure~\cite{sf02}.

\begin{figure}
\includegraphics[width=0.8\columnwidth]{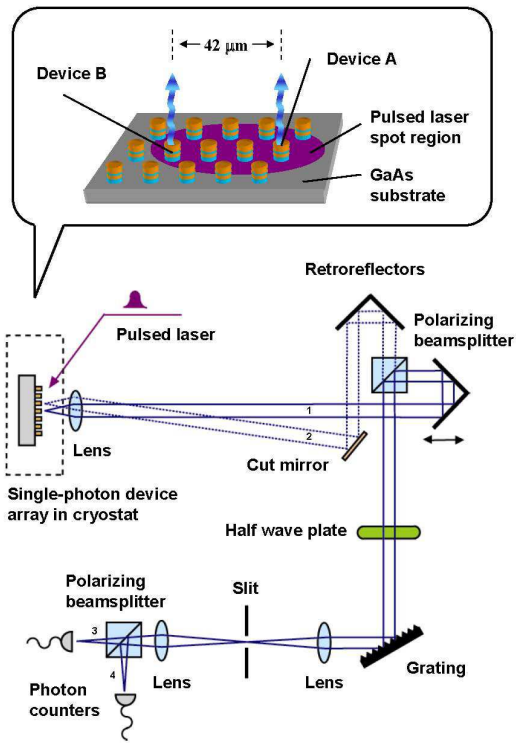}
\caption{\label{setup} (Color online) Experimental setup. The inset
shows the detailed configuration of the single-photon device matrix
array. The selected devices are separated by 42 $\mu$m, and the
same-pulsed pump laser beam spot excites the two devices. }
\end{figure}

The experimental setup of the two-photon interference experiment is
shown in Fig.~\ref{setup}. Many single-photon devices are fabricated
in a matrix array configuration with a pitch of 10~$\mu$m. The two particular devices A and B used in this
study are separated by a distance of 42~$\mu$m, which is within the
50-$\mu$m-diameter focus spot of the laser.  The device array is
cooled to 6~K in a coldfinger cryostat.  The photons are collected
by the same objective lens, but the light from one of the devices is
separated using a cut mirror placed close to a Michelson-type
interferometer. At the first polarizing beam splitter, the light
beams from the two devices are recombined into the same spatial
mode, but with orthogonal polarizations.  This mode undergoes
spectral filtering of a single emission line using a grating
monochromator.  When the half-wave plate is set to rotate the
polarization by 45$^\circ$, the two orthogonal incident
polarizations are equally mixed at the second polarizing beam
splitter, where quantum interference occurs.  If the photons are
identical, they will bunch into a single mode, reducing the
probability of a coincidence photon-count at the output.  The
outputs from this second polarizing beam splitter are detected using
a pair of single-photon-counting avalanche photodiodes. The
electronic signals from the photon counters are connected to a
high-resolution counting card in a start-stop configuration, which
generates a photon correlation histogram of the relative delay time
between the arrival times of the two photons on the detectors. This interferometer is different from a more typical apparatus using non-polarizing beam splitters, however this setup behaves equivalently while reducing the number of optical components for polarization and spectral filtering.

\begin{figure}
\includegraphics[width=0.75\columnwidth]{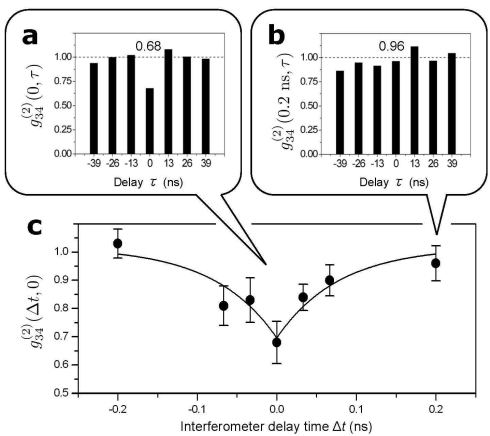}
\caption{\label{homfig} Two-photon correlation functions $g^{(2)}_{34}(\tau)$ from
the output of the interferometer. (a) Normalized photon correlation
histogram of the time delays of the arrival of two photons on the
photon counters, in the start-stop configuration, with no
interferometer path difference. The smallest normalized coincidence count at zero delay ($\sim$0.68) demonstrates two-photon interference. (b) Normalized photon correlation histogram, with 0.2~ns interferometer path difference. The normalized coincidence count at zero delay is higher ($\sim$0.96) than when the interferometer has no path difference.
(c) Normalized two-photon coincidence count on the photon counters,
plotted as a function of interferometer path-length difference.
Error bars are based on the Poisson fluctuation in the number
of counts on the uncorrected data. The solid curve is the fit
discussed in the text.}
\end{figure}

The experimental cross-correlation functions $g^{(2)}_{34}(\Delta t,\tau)$ (again with $\tau$ coarse-grained over the timing resolution of the detectors) for devices A and B at
two different interferometer delay times $\Delta t$ are shown in
Figs.~\ref{homfig}a and \ref{homfig}b.  These functions are calculated by integration of each peak and normalizing by the average intensity. When the setup is in a 50/50
beam-splitter configuration and the interferometer path delay $\Delta t$ is
zero, $g^{(2)}_{34}(0,0)$ reaches a minimum
of 0.68 with respect to the other peaks (Fig.~\ref{homfig}a). In contrast, when the path delay is 0.2~ns or larger, the two-photon coincidence probability $g^{(2)}_{34}(0.2~\text{ns},0)$ is close to one
(Fig.~\ref{homfig}b).  We observe that the minimum $g^{(2)}_{34}(\Delta t,0)$ is given when the interferometer delay time is zero with an accuracy of 3~ps.  The observed visibility is 31\%.

The HOM dip may be modeled in the following way.  When the source
successfully delivers a pair of photons from devices A and B, the
two-photon state can be written as
\begin{equation}
\ket\psi =\int ds \int dt \
    \alpha(t) \beta(t) \
    \tsc{a}{a}^\dag(t)\tsc{a}{b}^\dag(t) \ \ket{0},
\end{equation}
where $\tsc{a}{a}^\dag(t)$ and $\tsc{a}{b}^\dag(t)$  are the photon
creation operators for the two modes at time $t$, and  $\ket{0}$ is
the vacuum state.  The time dependent amplitude of the two photon
wavepackets are defined by the normalized complex-number functions
$\alpha(t)$ and $\beta(t)$ .  These functions model the spontaneous
emission and dephasing from each source. As a function of
interferometer path delay time, the normalized, coarse-grained coincidence count for the two detectors is given by~\cite{kiraz}
\begin{multline}
g^{(2)}_{34}(\Delta t,0) =
    \\ \frac{1}{2} \left[
    1-\text{Re}\int dt \int ds \
        \bigl\langle\alpha^*(s)
               \beta^*(t+\Delta t)
               \beta(s+\Delta t)
               \alpha(t)\bigr\rangle \right]
    \\
    +\ts{g}{back}.
\label{pceq}
\end{multline}
The last term $\ts{g}{back}$ is the background coincidence count, which includes non-zero $g^{(2)}(0)$  for devices A and B and extra stray light from other sources~\cite{kiraz}, and does not depend on $\Delta t$. The brackets $\langle\cdot\rangle$ average over many trials,
resulting in an overlap term phenomenologically given as
    $I\exp(-|\Delta t|/\ts\tau{c})$,
where $I$ is two-photon indistinguishability and $\ts\tau{c}$ is the
two-photon correlation time.  The indistinguishability $I$ is
upper-bounded by the ratio of the single-photon correlation time
divided by twice the radiative lifetime.  The correlation time $\ts\tau{c}$ is upper-bounded by the spontaneous lifetime~\cite{kiraz}. The observed indistingushability and correlation times are in general reduced from these upper bounds by frequency and timing jitter~\cite{legerog2}.

The indistinguishability and correlation time are found by fitting
Eq.~\ref{pceq} to the data in Fig.~\ref{homfig}c.  This fit
indicates that the excess background noise $\ts{g}{back}$ is about
$0.51 \pm 0.06 $.  This is higher than the amount of two-photon events that would be observed from two sources showing the $g^{2}(0)$ values indicated by Fig.~\ref{spectra}.   The reason for the discrepancy is that the amount of filtering is different in the two experiments.   The single-device experiments imaged each device onto a small pinhole to reduce background light from nearby devices to demonstrate single photon operation.  This spatial filtering is absent in the two-device experiment, increasing the number of two-photon events.  (The spectral and polarization filtering are unchanged between the two experiments).  The spurious extra photons in the two-device experiment are not expected to show any interference, and so the reduction of the two-photon probability is attributed to the interference of those photons originating from the spectrally matched devices A and B.

The fit indicates a correlation time of $\ts\tau{c} = 74
\pm$38~ps.  This time is comparable to the measured spontaneous
emission lifetimes of the devices A and B, suggesting nearly lifetime-limited sources; this suggests a potential for near-unity indistinguishability.  However, the fit indicates an indistinguishability of $I = 65 \pm 13 \%$, which is reduced from its ideal limit for several reasons.  First, a small wavelength mismatch or wavelength jitter between the two single photons may still exist, although it is unresolvable by our spectrometer.  The indistinguishability is also reduced by random timing jitter resulting in part from the noisy relaxation from higher-energy states to the lowest bound-exciton state from which we collect emission~\cite{sf02}. The width of this jitter is estimated to be 40~ps from the fitting to the streak-camera data in Fig.~\ref{spectra}, so this effect is small in comparison to  dephasing time. The larger contribution to the reduction of $I$ is probably the imperfect overlap of the spatial modes of the two photons.  In principle, this might be reduced by increased spatial filtering at the expense of the single-photon count-rate.

In conclusion, we have experimentally demonstrated the two-photon
quantum interference between indistinguishable photons
emitted by independent semiconductor devices with an observed
visibility of 31$\%$ and a deduced indistinguishability of 65$\%$.
Future directions for this work involve methods to improve the
indistinguishability.
One route for improvement is the fabrication of microcavities based on DBR or PC structures, which can enhance the spontaneous
emission rate and collection efficiency.  Further integration of such cavities with on-chip waveguides may enable compact, solid-state technologies for efficient, scalable optical quantum computation and quantum communication.

This work was supported by JST/SORST, NICT, and MURI (ARMY, DAAD 19-03-1-0199).

\newcommand{\etal}{{\textit{et al.}}}

\end{document}